\input harvmac.tex

\newread\epsffilein    
\newif\ifepsffileok    
\newif\ifepsfbbfound   
\newif\ifepsfverbose   
\newdimen\epsfxsize    
\newdimen\epsfysize    
\newdimen\epsftsize    
\newdimen\epsfrsize    
\newdimen\epsftmp      
\newdimen\pspoints     
\pspoints=1bp          
\epsfxsize=0pt         
\epsfysize=0pt         
\def\epsfbox#1{\global\def\epsfllx{72}\global\def\epsflly{72}%
   \global\def\epsfurx{540}\global\def\epsfury{720}%
   \def\lbracket{[}\def\testit{#1}\ifx\testit\lbracket
   \let\next=\epsfgetlitbb\else\let\next=\epsfnormal\fi\next{#1}}%
\def\epsfgetlitbb#1#2 #3 #4 #5]#6{\epsfgrab #2 #3 #4 #5 .\\%
   \epsfsetgraph{#6}}%
\def\epsfnormal#1{\epsfgetbb{#1}\epsfsetgraph{#1}}%
\def\epsfgetbb#1{%
%
%
\openin\epsffilein=#1
\ifeof\epsffilein\errmessage{I couldn't open #1, will ignore it}\else
%
%
   {\epsffileoktrue \chardef\other=12
    \def\do##1{\catcode`##1=\other}\dospecials \catcode`\ =10
    \loop
       \read\epsffilein to \epsffileline
       \ifeof\epsffilein\epsffileokfalse\else
%
%
          \expandafter\epsfaux\epsffileline:. \\%
       \fi
   \ifepsffileok\repeat
   \ifepsfbbfound\else
    \ifepsfverbose\message{No bounding box comment in #1; using defaults}\fi\fi
   }\closein\epsffilein\fi}%
%
%
\def\epsfclipstring{}
\def\epsfsetgraph#1{%
   \epsfrsize=\epsfury\pspoints
   \advance\epsfrsize by-\epsflly\pspoints
   \epsftsize=\epsfurx\pspoints
   \advance\epsftsize by-\epsfllx\pspoints
%
%
   \epsfxsize\epsfsize\epsftsize\epsfrsize
   \ifnum\epsfxsize=0 \ifnum\epsfysize=0
      \epsfxsize=\epsftsize \epsfysize=\epsfrsize
      \epsfrsize=0pt
%
%
     \else\epsftmp=\epsftsize \divide\epsftmp\epsfrsize
       \epsfxsize=\epsfysize \multiply\epsfxsize\epsftmp
       \multiply\epsftmp\epsfrsize \advance\epsftsize-\epsftmp
       \epsftmp=\epsfysize
       \loop \advance\epsftsize\epsftsize \divide\epsftmp 2
       \ifnum\epsftmp>0
          \ifnum\epsftsize<\epsfrsize\else
             \advance\epsftsize-\epsfrsize \advance\epsfxsize\epsftmp \fi
       \repeat
       \epsfrsize=0pt
     \fi
   \else \ifnum\epsfysize=0
     \epsftmp=\epsfrsize \divide\epsftmp\epsftsize
     \epsfysize=\epsfxsize \multiply\epsfysize\epsftmp   
     \multiply\epsftmp\epsftsize \advance\epsfrsize-\epsftmp
     \epsftmp=\epsfxsize
     \loop \advance\epsfrsize\epsfrsize \divide\epsftmp 2
     \ifnum\epsftmp>0
        \ifnum\epsfrsize<\epsftsize\else
           \advance\epsfrsize-\epsftsize \advance\epsfysize\epsftmp \fi
     \repeat
     \epsfrsize=0pt
    \else
     \epsfrsize=\epsfysize
    \fi
   \fi
%
%
   \ifepsfverbose\message{#1: width=\the\epsfxsize, height=\the\epsfysize}\fi
   \epsftmp=10\epsfxsize \divide\epsftmp\pspoints
   \vbox to\epsfysize{\vfil\hbox to\epsfxsize{%
      \ifnum\epsfrsize=0\relax
        \includegraphics{#1}%
      \else
        \epsfrsize=10\epsfysize \divide\epsfrsize\pspoints
        \includegraphics{#1}%
      \fi
      \hfil}}%
\global\epsfxsize=0pt\global\epsfysize=0pt}%
%
%
{\catcode`\%=12 \global\let\epsfpercent=
%
%
\long\def\epsfaux#1#2:#3\\{\ifx#1\epsfpercent
   \def\testit{#2}\ifx\testit\epsfbblit
      \epsfgrab #3 . . . \\%
      \epsffileokfalse
      \global\epsfbbfoundtrue
   \fi\else\ifx#1\par\else\epsffileokfalse\fi\fi}%
%
%
\def\epsfempty{}%
\def\epsfgrab #1 #2 #3 #4 #5\\{%
\global\def\epsfllx{#1}\ifx\epsfllx\epsfempty
      \epsfgrab #2 #3 #4 #5 .\\\else
   \global\def\epsflly{#2}%
   \global\def\epsfurx{#3}\global\def\epsfury{#4}\fi}%
%
%
\def\epsfsize#1#2{\epsfxsize}
%
%

%
\ifx\epsfbox\UnDeFiNeD\message{(NO epsf.tex, FIGURES WILL BE
IGNORED)}
\def\Fig.in#1{\vskip2in}
\else\message{(FIGURES WILL BE INCLUDED)}\def\Fig.in#1{#1}\fi
\def\iFig.#1#2#3{\xdef#1{Fig..~\the\Fig.no}
\goodbreak\topinsert\Fig.in{\centerline{#3}}%
\smallskip\centerline{\vbox{\baselineskip12pt
\advance\hsize by -1truein\noindent{\bf Fig..~\the\Fig.no:} #2}}
\bigskip\endinsert\global\advance\Fig.no by1}

\def\cst {{\rm const.}}

\def \ov {\over}

\def \lr { \lref}

\def\dd {\partial }

\def\l{\lambda}

\def \k {\kappa}

\def\n{\noindent}
\gdef \jnl#1, #2, #3, 1#4#5#6{ { #1~}{ #2} (1#4#5#6) #3}
\def\np {  Nucl.  Phys. }
\def \pl { Phys. Lett. }
\def \mpl { Mod. Phys. Lett. }
\def \prl { Phys. Rev. Lett. }
\def \pr  { Phys. Rev. }
\def \cqg { Class. Quant. Grav. }

\lr \cghs {C. Callan, S. Giddings, J. Harvey and A. Strominger,
\pr D45 (1992) R1005. }
\lr \hawk {S.W. Hawking, Commun. Math. Phys. 43 (1975) 199. }
\lr \rst {J.G. Russo, L. Susskind and L. Thorlacius,
\pr D46 (1992) 3444; \pr D47 (1993) 533.  }
\lr\payi {J. Park and P. Yi, \pl B317 (1993) 41.}
\lr \lohisc {D.J. Loranz, W.A. Hiscock and P.R. Anderson, \pr D52 (1995)
4554.} 
\lr\trivedi{S.P. Trivedi, \pr D47 (1993) 4233.}
\lr \gidstro{S.B. Giddings, {\it Quantum Mechanics of Black Holes},
Summer School in High Energy Physics and Cosmology, Trieste (1994), 
hep-th/9412138; A. Strominger, {\it Les Houches Lectures on Black Holes},
Les Houches Summer School (1994),
hep-th/9501071.}
\lr \fabru {A. Fabbri and J.G. Russo, \pr D53 (1996) 6995.}
\lr \balfab {R. Balbinot and A. Fabbri, gr-qc/9602047 (to appear in \cqg).}
\lr \fdu {W.G. Unruh, \pr\ D14 (1976) 870.}
\lr \bghs {B. Birnir, S.B. Giddings, J.A. Harvey, A. Strominger, 
\pr D46 (1992) 638.}
\lr \pyi {P. Yi, \prl\ 75 (1995) 382; \pr D53 (1996) 7041.}
\lr \fro {V.P. Frolov, \pr\ D46 (1992) 5383.}
\lr \napas {C.R. Nappi and A. Pasquinucci, \mpl A7 (1992) 3337.}
\lr \triv {S.P. Trivedi, \pr\ D47 (1993) 4233.}
\lr \anhilo {P.R. Anderson, W.A. Hiscock and D.J. Loranz, \prl 74 (1995) 
4365.} 
\lr \sutho {L. Susskind and L. Thorlacius, \np B382 (1992) 123.}
\lr \hawkelli {S.W. Hawking and G.F.R. Ellis, {\it The large scale
structure of space-time} 
(Cambridge University Press, Cambridge, England, 1973). }
\lr \wald {R. M. Wald, {\it General Relativity} (University of
Chicago Press, Chicago, 1984).} 
\lr \poimar {D. Markovic and E. Poisson, \prl\ 74 (1995) 1280.}
\lr \fabbri {A. Fabbri, {\it Classical stability of black hole 
Cauchy horizons in two-dimensional asymptotically flat space-times}, to 
appear} 
\lr \parentani {S. Massar and R. Parentani, hep-th/9603057; gr-qc/9603018.}
\baselineskip8pt
\Title{\vbox
{\baselineskip 6pt 
\hbox{SISSA-ISAS/100/96/EP} 
{\hbox{
   }}} }
{\vbox{\centerline { Two-dimensional black holes in accelerated frames: } 
\vskip .2in \centerline {quantum aspects}
}}
\bigskip\bigskip\bigskip
\vskip -20 true pt
\centerline  { {R. Balbinot  }}
 \smallskip \bigskip
\centerline{\it  Dipartimento di Fisica dell'Universit\`a di Bologna and
INFN sezione di Bologna,}
\smallskip
\centerline{\it  Via Irnerio 46, 40126 Bologna, Italy}
\smallskip
\centerline {\it   balbinot@bologna.infn.it}
\bigskip\bigskip\bigskip
\vskip -20 true pt
\centerline { A. Fabbri }
\smallskip \bigskip
\centerline {\it SISSA-ISAS and INFN sezione di Trieste,  }
\smallskip
\centerline {\it Via Beirut 2-4, 34014 Trieste, Italy}
\smallskip
\centerline {\it   fabbri@gandalf.sissa.it}
\bigskip\bigskip\bigskip
\bigskip\bigskip\bigskip
\centerline {\bf Abstract}
\bigskip
By considering charged black hole solutions of a one parameter family
of two dimensional dilaton gravity theories, one finds the existence
of quantum mechanically stable gravitational kinks with a simple
mass to charge relation. Unlike their Einsteinian counterpart (i.e.
extreme Reissner-Nordstr\"om), these have nonvanishing horizon surface
gravity.

\medskip
\baselineskip8pt
\noindent

\Date {July 1996}

\noblackbox
\baselineskip 14pt plus 2pt minus 2pt

\vfill\eject


\newsec{Introduction}

Hawking's discovery \hawk\ that semiclassically black holes radiate
has triggered a widespread debate concerning the ultimate fate of an 
evaporating black hole.
\par \n
According to one school of thought, black holes evaporate completely
in a finite time into the vacuum (\hawk, for an explicit two-dimensional
model see
for instance the RST model \rst).
Others advocate the existence of stable (zero temperature) remnants
of finite (probably Planckian) mass which should represent the
endpoint of the evaporation process (see for example the review
articles \gidstro).
\par \n
Connected to these two alternative issues is the question concerning
unitarity violations in the quantum mechanics of the black hole
formation-evaporation process.
\par
Zero temperature remnants are easily found in Einstein gravity
coupled to electromagnetism (or to some other abelian long range field).
This theory contains charged black hole solutions (Reissner-Nordstr\"om)
provided the mass $m_0$ of the hole equals or exceeds, in Planck
units, the conserved abelian charge $Q$.
\par \n
According to the conventional picture a black hole of mass
$m_0>|Q|$ will quantum mechanically evaporate until it reaches the
extremal value $m_0=|Q|$, at which point the Hawking temperature
vanishes and the evaporation ceases. Thus the extremal Reissner-Nordstr\"om
solutions are expected to be the endpoint of Hawking evaporation
and correspond to stable quantum ground states. \foot{We assume here
and in the following that there is no charged matter field
coupled to the abelian long range force. Therefore charged black holes
can not discharge themselves by spontaneous pair-creation.}
Similar results are found also in dilaton gravity theories coupled
to abelian fields (see for example Refs. \fro, \napas).
\par
The analysis of quantum fields propagating on these extreme black hole
spacetimes and their backreaction on the geometry has been 
extensively performed in the past years using two-dimensional 
approximation schemes (\triv\ and \payi). They revealed that despite
the stability against the evaporation process, the expectation
values $\langle T_{ab} \rangle$ of the energy-momentum tensor of
a massless scalar field, as measured by a free-falling observer,
is divergent on the horizon. \foot{Numerical evaluations of
$\langle T_{\mu\nu} \rangle$ in four spacetime dimensions
seem however not to reproduce this divergence (Ref. \anhilo), which
therefore should be considered just an artifact of the 
two-dimensional approximation scheme.}
Their backreaction on the geometry is, however, not problematic,
since in the `quantum corrected geometry' the metric and the 
scalar curvature $R$ appear to be well-behaved there.
\par
Quite different is the situation when no abelian field is present:
zero temperature remnants of finite mass with regular horizon
(gravitational kinks) simply do not exist. 
The `Boulware vacuum' construction of these kind of objects
produces unbounded $\langle T_{ab} \rangle$ at the black hole horizon.
This divergence is stronger than the one encountered in the previous
(extreme) case and as a result of the backreaction now a 
diverging scalar curvature $R$ is encountered as one approaches
the horizon (Refs. \bghs, \sutho).
\par
The purpose of this paper is an investigation of the existence
of kink like solutions in a 2d dilaton gravity theory recently 
proposed which generalizes the original CGHS theory.
In Ref. \fabru\ it has been shown the existence of a one parameter
theories all leading to the RST action \rst\ at the semiclassical level;
they are both classically and semiclassically exactly solvable.
They are described by the classical action
\eqn\accla{
S_n={1\ov{2\pi}}\int d^2 x \sqrt{-g}[ e^{-{2\ov n}\phi}
(R + {4\ov n}(\nabla\phi)^2) + 4\l^2 e^{-2\phi}],
}
where $R$ is the scalar curvature associated to the two-dimensional
metric tensor $g_{ab}$, $\phi$ is the dilaton field.
$n$ is the parameter labelling the different theories; $n=1$
reproduces the usual CGHS action \cghs.
\par
Static black hole solutions of these theories for $n\neq 1$
have the striking feature that the natural frame in which the
metric is static is not `asymptotically' minkowskian but Rindler like.
This, as we shall see, is the key feature to understand the existence
of our gravitational kinks.
\par
The introduction, in these theories, of an electromagnetic term
coupled to the dilaton and to the gravitational field breaks
as usual the conformal invariance of the theory, which is no longer
exactly solvable at the semiclassical level.
\par \n
However, as we shall see, classical exact solutions describing 
charged black holes can be found and also the corresponding semiclassical
corrections to these geometries can be worked out perturbatively.
\par \n
We shall show the conditions these solutions have to fulfil in order
to describe gravitational kinks.
The properties of these latter will be then compared to their 
extreme Reissner-Nordstr\"om like counterpart.

\newsec{Quantum fields in 2d Reissner-Nordstr\"om spacetime}

In this section we collect the most important results (for our 
purposes) concerning quantum effects in 2d Reissner-Nordstr\"om
spacetime.
\par \n
This spacetime is the unique solution of Einstein-Maxwell theory
which is spherically symmetric and asymptotically flat.
The two-dimensional section ($\theta=\cst$, $\phi=\cst$) of this
solution is described by the line element
\eqn\ago{
ds^2\equiv -e^{2\rho} dudv=-fdudv,
}
where the coordinates $u$ and $v$ are defined by
\eqn\nuco{
du=dt-{{dr}\ov{f}}, \ \ \ dv=dt+{{dr}\ov{f}},
}
and
\eqn\upa{
f=1-{{2m_0}\ov{r}}+{{Q^2}\ov{r^2}}
}
(we take $m_0>|Q|$). Eq. \ago\ describes a black hole of mass $m_0$ and
abelian charge $Q$. The coordinates ($u$, $v$) become, as $r\to\infty$,
the usual retarded and advanced time of special relativity.
\par \n
Associated to the staticity of these solutions is the existence
of a Killing vector $\dd_t$. The zeros of $f$ represent the
Killing horizons where the norm of $\dd_t$ vanishes. These null
surfaces are the event horizon at $r=r_+$ and the inner
(Cauchy) horizon at $r=r_-$, where
\eqn\evin{
r_{\pm}=m_0\pm \sqrt{m_0^2 - Q^2}.
}
We define also, for later use, the surface gravity $k$ at both
horizons
\eqn\hori{
k_{\pm}\equiv {1\ov 2} |\dd_r f|_{r_{\pm}}={{\sqrt{m_0^2 - 
Q^2}}\ov{r_{\pm}^2}}. }
The scalar curvature is simply
\eqn\cooo{
R=-f^{''}
}
and from eq. \upa\ we see that $r=0$ is the location of the singularity
and $r\to \infty$ defines the flat asymptotic region, where the
spacetime metric becomes Minkowski like. 
The Penrose diagram of this spacetime is represented (for $m_0>|Q|$)
in Fig. 1.
\par
When $m_0=|Q|$ the black hole is called extremal and in this case
\eqn\fest{
f=(1-{{m_0}\ov{r}})^2.
} 
This implies $r_+=r_-=m_0$ and $k_{\pm}=0$. The causal structure
of this solution is depicted in Fig. 2.
\par
The expectation values of the energy-momentum tensor of $N$ massless
scalar fields $f_i$ living on a two-dimensional spacetime can be
derived, knowing the conformal anomaly, simply by integrating
in the gauge of eq. \ago\ the conservation equations. The result
is 
\eqn\vevu{
\langle T_{uu} \rangle = - {N\ov{12\pi}}
(\dd_u\rho\dd_u\rho -\dd_u^2\rho - {{t_u(u)}\ov{8}}),
}
\eqn\vevd{
\langle T_{vv} \rangle = - {N\ov{12\pi}}
(\dd_v\rho\dd_v\rho - \dd_v^2\rho - {{t_v(v)}\ov{8}}),
}
\eqn\vevt{
\langle T^a_a \rangle = {{NR}\ov{24\pi}}.
}
Here units are chosen such that $\hbar=c=G=1$.
$t_u(u)$ and $t_v(v)$ are functions of their arguments and depend on
the choice of the quantum state in which the expectation values
have to be taken.
For our static space-time these are just constants to be fixed later
and
\eqn\vcxu{
\langle T_{uu} \rangle = {N\ov{96\pi}}(ff^{''}-{1\ov 2}f^{'2}+t_u),
}
\eqn\vcxd{
\langle T_{vv} \rangle = {N\ov{96\pi}}(ff^{''}-{1\ov 2}f^{'2}
+t_v),
}
\eqn\vcxt{
\langle T^a_a \rangle = - {{Nf^{''}}\ov{24\pi}},
}
where $f$ is given in eqs \upa\ or \fest.
\par
Because of the asymptotic minkowskian behaviour of the metric,
it seems reasonable to require that, as $r\to\infty$, $\langle T_{ab}
\rangle$ vanishes. This simply fixes the constants
\eqn\boul{
t_u=t_v=0.
}
This choice of state corresponds to the `Boulware vacuum', i.e. the state
that asymptotically corresponds to the usual vacuum state of Minkowski
quantum field theory.  
\par \n
However, when $m_0>|Q|$, this `natural choice' is not appropriate
to describe correctly the vacuum polarization induced by the fields
$f_i$ near the event horizon. In fact, in terms of a system of
coordinates ($U$, $V$) regular on the event horizon $r_+$
(Kruskal coordinates)
\eqn\krco{
U=-{{e^{-k_+ u}}\ov{k_+}}, \ \ \ V={{e^{k_+ v}}\ov{k_+}},
}
and vanishing respectively on the future and past sheet of this
surface, one has
\eqn\divi{
\langle T_{UU} \rangle \sim - {{k_+^2}\ov{U^2}}, \ \ \ 
\langle T_{VV} \rangle \sim - {{k_+^2}\ov{V^2}},
}
which are clearly divergent for $U\to 0$, $V\to 0$.
\par \n
The backreaction of this source on the geometry (Refs. \bghs, \sutho)
produces a diverging curvature on the event horizon
(the ADM mass stays however finite due to the asymptotic behaviour
of $\langle T_{ab} \rangle$ in the Boulware state).
\par \n
The presence of this singularity indicates that the Boulware state
can not be the relevant quantum state to describe vacuum polarization
in a black hole spacetime.
\par
Regularity of the stress tensor on both past and future event horizon
is obtained by the state (Hartle-Hawking) given by the choice
\eqn\har{
t_u=t_v=2k_+^2,
}
which ensures a sufficiently rapid vanishing of $\langle T_{uu}
\rangle = \langle T_{vv} \rangle$ on $r=r_+$.
In fact as $r\to r_+$
\eqn\tuvh{
\langle T_{uu} \rangle = \langle T_{vv} \rangle \sim
\cst (r-r_+)^2
}
which implies regularity of $\langle T_{ab} \rangle$ in the 
Kruskal frame \krco. \foot{This construction does not prevent
however the stress tensor to diverge on the inner (Cauchy)
horizon.}
However, as a result of this choice of state, asymptotically
\eqn\ashh{
\langle T_{uu} \rangle = \langle T_{vv} \rangle \to 
{{N k_+^2}\ov{48\pi}},
}
i.e. the stress tensor is no longer vanishing.
Physically the Hartle-Hawking state is a thermal state, describing
thermal equilibrium of a black hole and a heat bath at the Hawking
temperature $T_H={{k_+}\ov{2\pi}}$.\foot{Due to the behaviour in
eq. \ashh, the ADM mass of the geometry obtained by including 
the backreaction of $\langle T_{ab} \rangle$ is infinite.
Usually one eliminates this problem by enclosing the system in a box.}
\par
Summarizing the basic result, we have that as long as $m_0>|Q|$
the two conditions of vanishing of $\langle T_{ab} \rangle$ at infinity
and regularity on the event horizon cannot be fulfilled at the same time.
\par
The situation is significantly different in the extremal case $m_0=|Q|$.
Being now $r_+=r_-$ and $k_+=0$ the Boulware and the Hartle-Hawking
state become, in some sense, the same state.
Neverthless one can show (see for example Refs. \triv, \lohisc) that at the 
horizon a free falling observer will measure an energy density
\eqn\olpo{
\rho_{obs} \sim {{f^{'''}}\ov{f^{'}}}
}
which is divergent because $f$ has now a double zero at $r_+$ (see eq.
\fest).
\par \n
However, the divergence in eq. \olpo\ is in some sense weak, since the
geometry obtained by including the backreaction of this $\langle
T_{ab} \rangle$ (quantum corrected geometry) has regular curvature on
the horizon. \foot{However a mild
singularity appears in the second derivative of $R$ as seen by a
free falling observer.}
\par
We should stress the fact that here we focus our attention
to static black hole configurations, quantum-mechanically stable,
of finite mass and regular at the event horizon. These are natural
candidates for the final state of a black hole dynamically formed
by the collapse of matter which then evaporates. The only
candidate found is therefore the quantum version of the extremal
Reissner-Nordstr\"om black hole.
\par
Other examples with the same features are found in 2d dilaton gravity
theories like the charged extension of the CGHS model (see for
example Refs. \fro, \napas).

\newsec{Charged black holes in accelerated frames}

In Ref. \fabru\ a one parameter class of simple models of 2d
dilaton gravity theory was considered. The model is described
by the action \accla\
\eqn\accla{
S_n={1\ov{2\pi}}\int d^2 x \sqrt{-g}[ e^{-{2\ov n}\phi}
(R + {4\ov n}(\nabla\phi)^2) + 4\l^2 e^{-2\phi}].
}
The equations of motion derived from this action are
$$g_{\mu\nu} [{4\ov n} (-{1\ov 2} + {1\ov n})(\nabla \phi )^2- {2\ov n}
(\nabla^2 \phi )- 2\lambda^2 e^{{{2-2n}\ov{n}}\phi} ]+{4\ov n}
(1-{1\ov n})\partial_{\mu}\phi \partial_{\nu}\phi $$
\eqn\motoa{
+{2\ov n}\nabla_{\mu}\partial_{\nu}\phi =0\ ,
}
\eqn\motob{
{R\ov n}- {4\ov{n^2}}(\nabla\phi)^2 + {4\ov n}\nabla^2 \phi+
4\lambda^2 e^{{{2-2n}\ov{n}}\phi}=0\ .
}
As shown in appendix A, static solutions of these equations describing
black holes can be found.
\par
The general solution describing a static uncharged black hole can
be given in the `Schwarzschild-Rindler' gauge ($\sigma$, $t$) as
\eqn\schw{
ds^2=e^{2(1-n)\l\sigma}(-fdt^2+{1\ov f}d\sigma^2),
}
\eqn\schwu{
\phi=-n\l\sigma,
}
where $f$ is defined as
\eqn\fsch{
f=1- {{2m_0}\ov{\l}}e^{{2\ov n}\phi}.
}
Proceeding to the usual ADM construction, one finds that $2m_0$
represents the mass of the black hole.
\par \n
It is easy to realize that `asymptotically', i.e. as $\sigma\to\infty$,
the metric in eq. \schw\ becomes flat but it is expressed in 
Rindler coordinates instead of the usual Minkowski ones.
\par
Adding to the action eq. \accla\ an electromagnetic term of the form
\eqn\azem{
S_{EM}={1\ov{2\pi}}\int d^2 \sqrt{-g}e^{-2a\phi}(-2F_{\mu\nu}F^{\mu\nu}),
}
where $a={{2-n}\ov n}$ and $F_{\mu\nu}$ is the electromagnetic tensor,
one can find solutions describing black holes which carry an abelian
charge $Q$.
The equations of motion for $F_{\mu\nu}$ derived from eq. \azem\
are
\eqn\eqem{
\nabla_{\nu} (e^{-2a\phi}F^{\mu\nu})=0,
}
which are easily integrated leading to
\eqn\efem{
F_{\mu\nu}=Qe^{2a\phi}e_{\mu\nu},
}
where $e_{\mu\nu}=e_{[\mu\nu]}$ and $e_{01}=\sqrt{-g}$. \foot{Note that
for $n=2$ the electromagnetic field is completely decoupled from
the dilaton and the insertion of $S_{EM}$, once eqs. \efem\ have
been taken into account, is then equivalent to the addition of 
a 2d cosmological constant to the action.}
\par \n
The solution of the field equations for the metric and the dilaton
are again given by eqs. \schw\ and \schwu\ but where $f$ now is
\eqn\frn{
f=1-{{2m_0}\ov{\lambda}}e^{{2\ov n}\phi}+
{{Q^2}\ov{\lambda^2}}e^{{4\ov n}\phi}.
}
This solution describes static black holes of mass $m_0$ and abelian
charge $Q$. 
\par
The causal structure of these solutions can be studied following the
scheme of Ref. \balfab.
The line element is transformed to the chiral form
\eqn\chiral{
ds^2=-h(v,x)dv^2 +2dvdx.
}
by the coordinate transformation
\eqn\tra{
x = {{e^{2(1-n)\l\sigma}}\ov{2(1-n)\l}},\ \ \ v=t+ \int {{d\sigma}\ov{f}},
}
which yields
\eqn\cupo{
h(v,x)=2(1-n)\l x - {{2m_0}\ov{\l}}[2(1-n)\l x]^{{n}\ov{n-1}} +
{{Q^2}\ov{\l^2}}[2(1-n)\l x]^{{n+1}\ov{n-1}}.
}
With this choice of coordinates $v$ labels ingoing null lines and $x$
is an appropriate normalized affine parameter on them. Starting from
the form eq. \chiral\ of the metric one can construct the fundamental
building blocks and the resulting Penrose diagrams.
\par \n
For the purposes of the present paper we need just to outline
some basic characteristics of the spacetimes.
\par \n
For $n<1$ and $m_0>|Q|$ the solution presents two horizons located
at $\sigma=\sigma_{\pm}$, where 
\eqn\hori{
e^{2\lambda\sigma_{\pm}}=e^{-{2\ov n}\phi_{\pm}}=
{{(m_0\pm \sqrt{m_0^2-Q^2})}\ov{\lambda}}\ .
}
$\sigma_+$ is the location of the outer (event) horizon, and $\sigma_-$
is the inner (Cauchy) horizon. For $m_0=|Q|$ the two horizons coalesce.
The essential singularity is at $\sigma=-\infty$, whereas the
asymptotically flat region is at $\sigma=+\infty$.
\par
Our solution, apart a conformal factor, is the same as the charged
(Reissner-Nordstr\"om like) solution of the CGHS ($n=1$) model.
The corresponding Penrose (conformal) diagrams are therefore the
same as those represented in Figs. 1 and 2.
\par \n
The presence of the conformal factor makes however the metric to
approach asymptotically a Rindler like form instead of the usual
Minkowski one. One can introduce minkowskian null coordinates
$y^{\pm}$ defined by 
\eqn\mink{
\l y^+ = {{e^{\l (1-n) v}}\ov{(1-n)}},
}
\eqn\minku{
-\l y^- = {{e^{-\l (1-n)u}}\ov{(1-n)}},
}
where $u=t-\int {{d\sigma}\ov f}$ and $v=t+\int {{d\sigma}\ov f}$,
but the metric is no longer static in this frame. It takes the form
(for $Q=0$)
\eqn\memi{
ds^2  
= -{{dy^+dy^-}\ov{\big[ 1+{{2m_0}\ov{\l}}\big( -\l^2 (1-n)^2 y^+y^- 
\big) ^{{1\ov{n-1}}} \big] ^n }}
\  .
}
\par
For $n>1$ one has the same features described above except for the
fact that $\sigma=+\infty$ does not correspond anymore to true
infinity, since inertial observers can reach this region in a finite
proper time. One can show that the null surface $\sigma=+\infty$ 
is an acceleration horizon for $\sigma=\cst$ observers.
\par
A useful quantity in the description of horizons is the local
surface gravity $k$, defined by (see for example \wald)
\eqn\surgra{
\k={1\ov 2}\big| {{\dd_{\sigma} g_{tt}}
\ov{\sqrt{-g_{\sigma\sigma}g_{tt}}}} \big| =
\big| \l (1-n) f + {1\ov 2} f_{,\sigma} \big|\ .
}
One finds that at the black hole horizons $\sigma_{\pm}$ 
\eqn\sagra{
\k_{\pm} = {1\ov 2}|f_{,\sigma_{\pm}}|
= \l e^{-2\l\sigma_{\pm}} (e^{2\l\sigma_+}
 - e^{2\l\sigma_- }).
}
Note that, provided $m_0>|Q|$, it is always $k_->k_+$ and that
$k_+=k_-=0$ for the extreme $m_0=|Q|$ hole.
\par \n
In the case $n>1$ one can also define the surface gravity
at the acceleration horizon
\eqn\suac{
k_{ah}\equiv k(\sigma=+\infty)=\l (n-1).
}
\par
Since in what follows we are interested in solutions exhibiting
an asymptotic region, only the case $n<1$ will be considered.

\newsec{Quantum kinks}

The behaviour of a conformally invariant quantum field propagating
on the charged black hole solution of our 2d dilaton gravity action
eqs. \accla\ and \azem\ is analysed within the framework developed
in section 2. We first write the metric in a conformal flat form
\eqn\cfga{
ds^2\equiv -e^{2\rho}dudv= -Ffdudv,
}
where $F=e^{2(1-n)\l\sigma}$ and $f$ is defined in eq. \frn.
$u$ and $v$ are usual advanced and retarded null coordinates
$dv=dt+{{d\sigma}\ov f}$, $du=dt-{{d\sigma}\ov f}$.
\par \n
The stress tensor $\langle T_{ab} \rangle$ is given as before by
eqs. \vevu, \vevd\ and \vevt. Again for static equilibrium
configurations we require that $t_u=t_v=C=\cst$.
The value of this constant reflects the choice of quantum state
in which to evaluate the expectation values.
\par
Our choice is to consider as `physical' only those configurations
which have regular event horizon, i.e. Hartle-Hawking state.
The quantum field operator should therefore be expanded
in normal modes
\eqn\modi{
e^{-iwx^+}, \ \ \ e^{-iwx^-}
}
where $x^{\pm}$ are Kruskal like coordinates, regular on the event horizon.
These are related to our `Schwarzschild-Rindler' frame by the following
coordinate transformations
\eqn\klas{
x^+={{e^{k_+ v}}\ov{k_+}}, \ \ \ x^-=- {{e^{-k_+ u}}\ov{k_+}}.
}
The value of $C$ is therefore simply the schwarzian derivative between
the Kruskal and the Schwarzschild-Rindler sets.
The end result is that the stress tensor $\langle T_{ab} \rangle$
evaluated in the Hartle-Hawking state $|H\rangle$ is
\eqn\eqcoa{
\langle H|T_{uu}|H\rangle =\langle H|T_{vv}|H\rangle =
{N\ov{96\pi}}[ff_{,\sigma\sigma}-{1\ov 2}f_{,\sigma}^2 -
2(1-n)^2\l^2 f^2 + 2k_+^2],
}
\eqn\eqcob{
\langle H|T^a_a|H\rangle = {{NR}\ov{24\pi}}=
-{N\ov{24\pi}}e^{-2(1-n)\l\sigma}[f_{,\sigma\sigma}+2(1-n)\l f_{,\sigma}].
}
These expressions can be checked to be regular both on the future and
the past sheet of the event horizon.
\par
Note that in general the Kruskal coordinates $x^{\pm}$ are not
inertial coordinates at infinity and therefore $\langle H|
T_{ab}|H\rangle$ does not vanish asymptotically, namely
as $\sigma\to +\infty$
\eqn\emev{
\langle H|T_{uu}|H\rangle = \langle H| T_{vv}|H \rangle
\simeq  {N\ov{48\pi}}
[k_+^2 - \l^2 (1-n)^2]
}
and in terms of the Minkowski coordinates at infinity $y^{\pm}$,
defined in eqs. \mink\ and \minku, one has
\eqn\toro{
\langle H|T_{y^{\pm}y^{\pm}}|H\rangle \simeq
{{N}\ov{48\pi (1-n)^2 \l^2 y^{\pm 2}}}[k_+^2 - \l^2 (1-n)^2 ],
}
\eqn\torob{
x^{\pm}=\pm{1\ov{k_+}}[\pm (1-n)\l y^{\pm}]^{{{k_+}\ov{\l (1-n)}}}.
}
However when the surface gravity $k_+$ of the hole satisfies 
\eqn\quin{
k_+ = \l (1-n)
}
we see that eq. \toro\ vanishes and the relation eq. \torob\
becomes trivial, i.e. $x^{\pm}=y^{\pm}$. This is the basic result
of our paper.
\par \n
From the definition of $k_+$ (eq. \sagra) one can rewrite the
condition eq. \quin\ as
\eqn\recla{
Q^2={{4n}\ov{(1+n)^2}}m_0^2,
}
which implies $m_0>|Q|$ for $n\neq 1$ ($n$ positive).
\par
What we have just shown is a remarkable property of our charged
black hole solution: provided that the mass and the charge of the hole
satisfy eq. \recla, the quantum state $|H\rangle$ yields a stress
tensor regular on the event horizon and vanishing asymptotically.
These black holes are therefore natural candidates for gravitational
kinks.
\par \n
Being $k_+\neq 0$ the event and inner horizons of the black hole
are distinct and located respectively at
\eqn\caucci{
e^{2\l\sigma_+}={2\ov{(1+n)}}{{m_0}\ov{\l}},\ \ \ e^{2\l\sigma_-}=
{{2n}\ov{(1+n)}}{{m_0}\ov{\l}}.
}
\par
The backreaction of the stress tensor on the geometry can be 
evaluated perturbatively (see appendix B for the details) yielding
for the quantum corrected geometry of the kink (we reinsert 
$\hbar$ in the formulas)
\eqn\fqua{
f(\sigma)\sim 1-{{2m_0}\ov{\l}}e^{-2\l\sigma}+
 [{{4n}\ov{(1+n)^2}}{{m_0^2}\ov{\l^2}}-\hbar {{Nm_0}\ov {9\pi\l}}n
 (n^2 -2n+3)]e^{-4\l\sigma} + O(e^{-6\l\sigma}),
}
which shows that the ADM mass of the quantum kink is still $2m_0$.
\par \n
For the dilaton field we find
\eqn\quadi{
e^{-{{\phi}\ov{n}}}\sim e^{\l\sigma}(1+\hbar 
 {{Nm_0}\ov{{18\pi\l}}}
 (2n-n^2 ) e^{-4\l\sigma}).
}
In appendix B we also give the corrected value of the
event horizon $\bar \sigma_+$. Both metric and dilaton field
are regular there. 
\par 
A very different behaviour is observed
near the Cauchy horizon $\sigma=\sigma_-$.  
$\langle T_{ab} \rangle$ diverges badly there. The quantum corrected
geometry must be very different from the classical one. Following the
argument given in Ref. \fabbri\ one expects the curvature to diverge
like ${1\ov {x^+ (-\ln(-k_- x^+))^{2-n}}}$ as $x^+\to 0$ ($x^+$ is now the 
Kruskal advanced null coordinate on  the Cauchy horizon).

\newsec{Conclusion}

In this paper we have found the existence of charged black hole
solutions of a one parameter ($n$) 2d dilaton gravity action
which generalizes the CGHS action.
This latter corresponds to the value $n=1$.
\par
At the classical level the static black holes of charge $Q$
do not seem to differ drastically from the usual Reissner-Nordstr\"om
like solutions of General Relativity.
They have two horizons (outer and inner) provided that 
$m_0>|Q|$. For $m_0=|Q|$ the two horizons coalesce
and we have an extreme black hole with zero surface gravity
on the horizon. 
\par \n
However the presence of the conformal factor
(which is non trivial for $n\neq 1$) in the expression of the metric eq.
\schw\ makes the physics of these objects more subtle. 
These black holes are in fact static only when viewed by a Rindler  
like frame. Asymptotically to this frame one can associate an acceleration,
with respect to an inertial (Minkowsky) frame,
\eqn\coac{
a_M=\l (1-n).
}
So one can think our black holes to be accelerated with this
acceleration. 
\par
Focusing our attention to semiclassical effects, we looked for 
configurations which might represent stable ground states for black
hole evaporation. To this end we searched for static, finite mass 
black hole solutions with regular event horizon and vanishing 
quantum radiation at infinity.
These configurations (gravitational kinks) indeed exist provided
the surface gravity $k_+$ of the event horizon equals the above
acceleration $a_M$
\eqn\coeq{
k_+=a_M.
}
This relation constraints mass and charge of the kink to fulfil
the following equation
\eqn\mcco{
Q^2={{4n}\ov{(1+n)^2}}m_0^2.
}
\par \n
From eq. \coeq\ we see that the surface gravity of the event horizon
of these black holes is in general nonvanishing (except for $n=1$),
i.e. the event horizon is not degenerate. Furthemore from eq. \mcco\
we see that the mass of these states can be made arbitrarily large
compared to $Q$ by lowering $n$ towards zero. 
Note that eq. \mcco\ requires $m_0>|Q|$ and therefore our extremal
solution $m_0=|Q|$ is not a kink.
\par 
These characterizing
features distinguish substantially our gravitational kinks from the
extreme Reissner-Nordstr\"om black hole. 
\par \n
For this latter
mass and charge are related by $m_0=|Q|$.
Furthemore, the horizon surface gravity vanishes (degenerate horizon),
which makes the quantum stress tensor to diverge on the degenerate
horizon. This divergence, however, causes no effect
on the spacetime geometry once the backreaction is properly taken
into account.
\par
How do we understand the relation eq. \coeq\ which leads to the 
gravitational kink configurations described by eq. \mcco?
\par \n
We think the situation presented here resembles the discussion
of Yi \pyi\ on the quantum stability of accelerated black holes 
based on the Ernst metric. 
\par \n
This metric represents two opposite
magnetically charged Reissner-Nordstr\"om black holes uniformly
accelerated away from each other, where the driving force is an
external magnetic field. 
\par \n
According to \pyi\ stability under thermal evaporation
is obtained by imposing the acceleration temperature 
(associated to the acceleration horizon of the Ernst metric)
to be equal
to the Hawking temperature $T_H$ (associated to the event horizon). 
\par \n
As a consequence, it is 
shown that the relation between regular coordinates at the event and
at the acceleration horizons is of the type $x^{\pm}=-{1\ov{y^{\pm}}}$,
making the Bogolubov transformation between 
the two basis trivial. \foot{The same relation exists in the case of an 
extremal 
Reissner-Nordstr\"om black hole between the Kruskal coordinates regular at 
the 
horizon and the asymptotic inertial ones.}
\par \n
This result has, however, been criticized by Massar and Parentani \parentani,
who emphasize the crucial role played by the presence of a second black hole.
According to the above authors, the decoherence of the two black holes, 
the independent spread of their masses around the mean and any breaking
of the exact boost invariance of the Ernst metric will lead inevitably
to the emission of a steady flux of particles by the holes. 
\par \n
Returning to our work, 
some fundamental differences should be stressed. In our construction
there is just one black hole and 
no acceleration horizon is present; this latter is 
replaced by null infinity. 
Furthemore, inertial coordinates there ($y^{\pm}$)
and regular coordinates on the event horizon ($x^{\pm}$)
simply coincide: $x^{\pm}=y^{\pm}$.

\bigskip \bigskip
\noindent $\underline {\rm Acknowledgements}$: We thank D. Amati
and R. Parentani for useful discussions.

\appendix {A} {Classical Solutions} 

In this appendix we develop a method for solving the equations of motion
deriving from the action $S=S_n + S_{EM}+S_M$, where $S_n$ is given in eq. 
\accla, $S_{EM}$ in eq. \azem\ and $S_M$ is some matter source.
We will consider only static frames and metric of the
 form 
\eqn\apml{
ds^2=e^{2(1-n)\l\sigma}(-f(\sigma)dt^2 + {{d\sigma^2}\ov{f(\sigma)}}),
}
where $f(\sigma)$ is an arbitrary function of its argument
to be determined by the field equations.
The other unknown function is the
dilaton. 
It turns out, for our purposes,
that a useful parametrization of $g_{\mu\nu}$ and $\phi$
is given by the following equations
\eqn\dipi{
{{e^{-{1\ov n}\phi}}\ov{\l}}=\int d\sigma e^{2(1-n)\l\sigma -\l g(\sigma)},
\ \ \
f(\sigma)=e^{2\l g(\sigma) -2(1-n)\l\sigma} h(\sigma),
}
where 
\eqn\cvbn{
h(\sigma)=e^{-2\phi}(1-{{2m(\sigma)}\ov{\l}}e^{{2\ov n}\phi}+{{Q^2}\ov{\l^2}}
e^{{4\ov n}\phi}).
}
\par
In terms of the two functions $g(\sigma)$ and $m(\sigma)$ the equations
\motoa\ can be rewritten in the form
\eqn\chjk{
\dd_{\sigma}m = - {{T^{t}_t}\ov 2}e^{(2-{1\ov n})\phi + 2(1-n)\l\sigma -\l 
g(\sigma)}
}
and
\eqn\cmlw{
\dd_{\sigma}g = (1-2n) + {{(T^{\sigma}_{\sigma}-T^t_t)}\ov{2\l^2 h(\sigma)}}
e^{2(1-n)\l\sigma -\l g(\sigma)+{1\ov n}\phi},
}
where $T_{ab}$ is the stress tensor associated to the source $S_M$.
\par
In the vacuum, $T_{ab}=0$, we get simply
\eqn\vacj{
m=m_0, \ \ \ g(\sigma)=(1-2n)\sigma
}
and therefore the solution of the equations of motion is
given by 
\eqn\poiu{
f_{cl}=1-{{2m_0}\ov{\lambda}}e^{-2\l\sigma}+
{{Q^2}\ov{\lambda^2}}e^{-4\l\sigma}, \ \ \ \phi=-n\l\sigma.
}
 
\appendix {B} {Quantum corrected solutions}

We now turn to the problem of calculating the quantum corrections to the 
 solution
given in eq. \poiu.
The difference with respect to the classical case is that in eqs.
\chjk\ and \cmlw\  the quantum energy-momentum $\langle T_{ab} \rangle$
must be inserted as a source. 
\par \n
We write down the following decomposition
\eqn\cccc{
m=m_0+m_q (\sigma), \ \ \ g(\sigma)=(1-2n)\sigma +g_q(\sigma)
}
where $m_q$ and $g_q$ are of order $\hbar$.
\par \n
According to eqs. \chjk\ and \cmlw\ we get, to the same order, 
\eqn\fgtt{
\dd_{\sigma}m_q(\sigma)={{\langle T_{tt}\rangle}\ov{2f_{cl}(\sigma)}}
}
and
\eqn\qefe{
\dd_{\sigma}g_q(\sigma)=
{{e^{-2\l\sigma}}\ov{2\l^2 f_{cl}^2(\sigma)}}(f_{cl}^2(\sigma)
\langle T_{\sigma\sigma}\rangle +\langle T_{tt} \rangle ),
}
where $\langle T_{ab} \rangle$ are calculated on the classical 
background metric.
\par
Let us consider first eq. \fgtt.
The source is
\eqn\tuvv{
\langle T_{tt} \rangle = 4(\langle T_{vv}\rangle +\langle T_{uu}\rangle +
2\langle T_{uv}\rangle),
}
where (see eqs. \eqcoa\ and \eqcob\ and hereafter $f\equiv f_{cl}$)
\eqn\tuuu{
\langle T_{uu}\rangle =\langle T_{vv}\rangle =
{{N\hbar }\ov{48\pi}}[ (1-n)^2 \l^2 (1-f^2) - {1\ov 4}f_{,\sigma}^2 + 
{1\ov 2} f f_{,\sigma \sigma}]
}
and
\eqn\vvvv{
\langle T_{uv} \rangle = {{N\hbar}\ov{48\pi}}
[{1\ov 2} f f_{,\sigma \sigma}+ \l (1-n) f f_{,\sigma}].
}
Inserting it we obtain
\eqn\maqu{
\dd_{\sigma}m_{q}(\sigma)={{N\hbar}\ov{12\pi}}[f_{,\sigma\sigma}+ \l 
(1-n)f_{,\sigma} + {{(1-n)^2 \l^2 (1-f^2) -{1\ov 4}f_{,\sigma}^2}\ov{f}}]
}
and after some algebra we finally find 
$$ m(\sigma)=m_0 + {{\hbar N}\ov{6\pi}}\{  m_0[(1+n)-{{(1-n)^2}\ov{2}}]
e^{-2\l\sigma}
 +{{m_0^2}\ov{2\l}} {{n(n-7)}\ov{(1+n)}}e^{-4\l\sigma} + 
$$
\eqn\mfin{ + \l {{(1+n)(1-n)^2}\ov{4n}}
\ln |{{e^{-2\l\sigma} - {{(1+n)\l}\ov{2nm_0}}}\ov{{{(1+n)\l}\ov{2nm_0}}}}|
 \} ,
 }
where the kink relation $Q^2={{4n}\ov{(1+n)^2}}m_0^2$ was used.
\par
For the calculation of $g_q(\sigma)$ we proceed similarly and from eq. \qefe\
we get
\eqn\gqua{
\dd_{\sigma}g_q(\sigma)={{4e^{-2\l\sigma}}\ov{\l^2}}{{\langle 
T_{uu}\rangle + 
 \langle T_{vv}\rangle}\ov{f^2}}=
 {{\hbar N}\ov{6\pi}}{{e^{-2\l\sigma}\ov{\l^2}}}
 [{{(1-n)^2 \l^2 (1-f^2)-{1\ov 4}f_{,\sigma}^2 + {1\ov 2}f f_{,\sigma\sigma}
 }\ov{{f^2}}}].
 }
 We obtain
$$ g(\sigma)= (1-2n)\sigma + {{\hbar N}\ov{6\pi}}[ 
 {{(n^2 - 2n -3)}\ov{2\l}} e^{-2\l\sigma}+
 {1\ov{m_0}} {{(1+n)^2(n-1)}\ov{4n}}
 {{e^{-2\l\sigma}}\ov{(e^{-2\l\sigma}- {{(1+n)\l}\ov{2nm_0}})}}
 -
 $$
  \eqn\gggg{-{1\ov {m_0}}{{(1+n)^2}\ov{2n}}
  \ln |{{e^{-2\l\sigma} - {{(1+n)\l}\ov{2nm_0}}}\ov{{{(1+n)\l}\ov{2nm_0}}}}|
  \ ]
  .
  }
  \par
  We can use these results to determine the quantum-corrected 
radius 
  of the event horizon, which 
  is located at $\sigma=\bar \sigma^+$ where
  \eqn\quev{
  e^{-{2\ov n}\phi(\bar \sigma^+)}={{2 m(\bar \sigma^+)}\ov{\l (1+n)}},
  }
that is
$$ e^{2\l\bar\sigma^+}={{2m_0}\ov{\l (1+n)}}+ 
{{\hbar N}\ov{6\pi}}\{ {{(1+n)(-5n+14)}\ov{4}}+
{{(n^2-6n-3)}\ov{2n}}\ln (1-n) +$$
\eqn\qceh{
 + {{(1+n)(5-n)}\ov{2\sqrt{n}}}
\ln {{1-\sqrt{n}}\ov{1+\sqrt{n}}} \} .
}
  \par
  Note finally that close to the Cauchy horizon $\sigma\sim\sigma^-$ 
  the perturbative
  terms $\sim O(\hbar)$ are huge and diverge at $\sigma^-$. It signals
  that in this region our perturbative approach breaks down.
\vfill \eject
\listrefs 
   
\vfill \eject
{
 \epsfxsize=6cm \epsfysize=8cm 
 \epsfbox{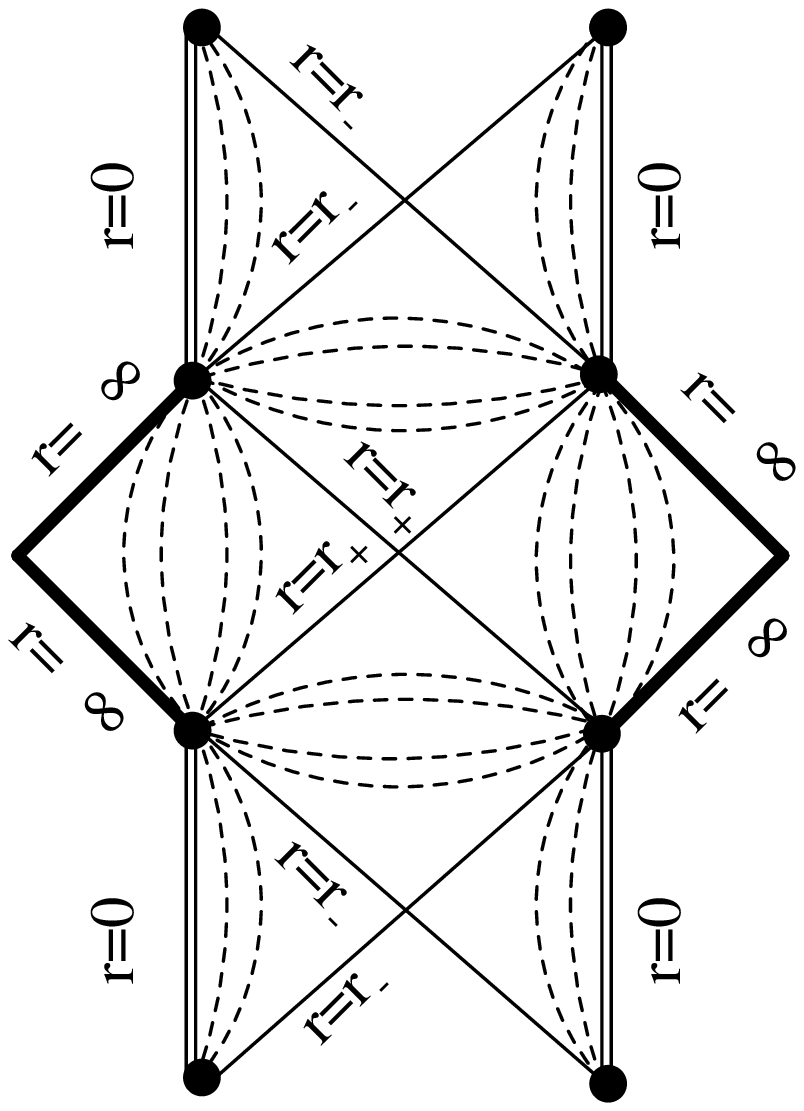}
 }
{\bf {Fig. 1:}}{
Penrose diagram of the Reissner-Nordstr\"om spacetime for $m_0>|Q|$. 
Double lines represent the singularity, 
dashed lines the curves $r=\cst$, regular lines the horizons and 
thick lines the asymptotic region.}

{
\epsfxsize=4cm \epsfysize=8cm 
\epsfbox{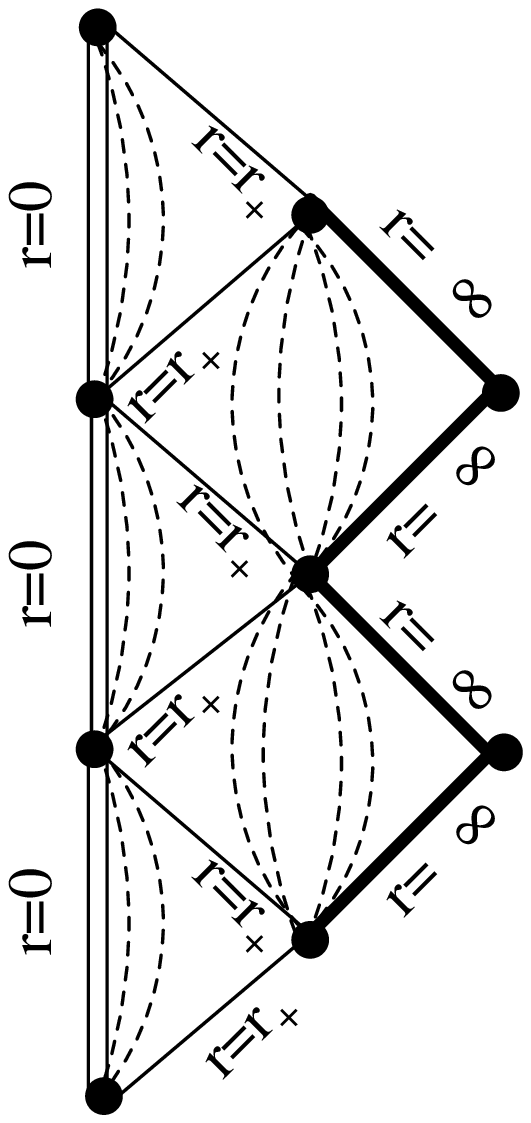}}

{\bf {Fig. 2:}}{ 
Causal structure of the extremal Reissner-Nordstr\"om geometry ($m_0=|Q|$).}

\vfill\eject

\end